\documentclass[a4,12pt,reqno]{article}
\usepackage{amsmath,amssymb,graphics,color}
\usepackage{fixmath} %This package is used to bold math symbols, works for both Greek and Latin letters using command \mathbold.
\usepackage[margin=2.5cm]{geometry}

\begin{document}
\title{Spherically Symmetric Cosmological Model with Charged Anisotropic Fluid in Higher Dimensional Rosen's Bimetric Theory of Gravitation}
\author{A.H. Hasmani$^1$ and D.N. Pandya$^2$\\\small{Department of Mathematics, Sardar Patel University, Vallabh Vidyanagar}}
\thanks{ah\_hasmani@spuvvn.edu$^1$, dee6788@gmail.com$^2$}
\date{} %TO remove date when "article" class is used
{\let\newpage\relax\maketitle}
\noindent \textbf{Abstract} : In this paper we have obtained cosmological models for the static spherically symmetric spacetime with charged anisotropic fluid distribution in $(n+2)$-dimensions in context of Rosen's Bimetric General Relativity (BGR). An exact solution is obtained and a special case is considered.
\section{Introduction}
The understanding of gravitation is a problem of concern to many researchers. Many theories are available for the explanation. Among the successful theories Newton's theory (in limited sense) and Einstein's theory are leading. However, new theories either due to modification or with new ideas are continuously proposed. Rosen (see \cite{Rosen1980}) formulated a new theory of gravitation called the Bimetric General Relativity (BGR). In this theory gravity is attributed to a curved spacetime described by the metric,
\begin{equation}
ds^2 = g_{ij} dx_i dx_j,
\end{equation}
a second metric tensor in the background space is described by
\begin{equation}
d\sigma^2 = \gamma_{ij} dx_i dx_j.
\end{equation}
The field equations in BGR are written in the form of Einstein's field equations, but with an additional term on the right hand side,
\begin{equation}\label{EFE}
G_{\nu}^{\mu} = - 8 \pi T_{\nu}^{\mu} + S_{\nu}^{\mu},
\end{equation}
where $G_{\nu}^{\mu}$ is the Einstein tensor, $T_{\nu}^{\mu}$ energy-momentum tensor of matter distribution and
\begin{equation}
S_\nu^\mu = \frac{3}{a^2}(\gamma_{\nu\alpha}g^{\alpha\mu}-\frac{1}{2}\delta_\nu^\mu g^{\alpha\beta}\gamma_{\alpha\beta}),
\end{equation}
where $a$ is a constant scale parameter. The order of this scale parameter is related to the size of the universe.\\
\indent In this paper we present an exact solution of the field equations for charged anisotropic fluid in $(n+2)$-dimensional BGR proposed by Rosen (see \cite{Rosen1980}). We have followed the method developed by Khadekar and Kandalkar (see \cite{Khadekarkandalkar2004}) by introducing the \emph{generating function} $G(r)$ which determines the relevant physical variables as well as the metrical coefficient and a function $w(r)$ measuring the degree of anisotropy, this function is called \emph{anisotropic function}. The General Relativity analogue of the charged anisotropic fluid in 4-dimensions was considered by Singh et al. (see \cite{SinghHelmi1995}) and results obtained here match with those of obtained there. Also in absence of charge, results obtained in this paper match with the one obtained by Kandalkar and Gawande (see \cite{kandalkarGawande2008}) for the case of Higher Dimensional General Theory of Relativity. Moreover for $n=2$, results in this paper matches with the results obtained by Hasmani and Pandya (see \cite{HasmaniDisha}) for 4-dimensional anisotropic charged matter in BGR.

\section{Metrics and Field Equations}
The general static spherically symmetric line element may be expressed as
\begin{equation}\label{metric}
ds^{2} = - e^{\lambda(r)}dr^2 - r^2 d\Omega^2 + e^{\nu(r)} dt^2,
\end{equation}
where
\begin{equation}
d\Omega^2 = d\theta_1^2 + \sin^2\theta_1 d\theta_2^2 + \sin^2\theta_1 \sin^2\theta_2 d\theta_3^2 + \cdot \cdot  \cdot+ \bigg[\prod_{i=1}^{n-1} \sin^2\theta_i\bigg] d\theta_n^2.
\end{equation}
Consider the background flat metric $\gamma_{\mu \nu}$ in $(n+2)$-dimensional analogue of static de-Sitter form as
\begin{equation}
d\sigma^2 = -\bigg(1-\frac{r^2}{a^2}\bigg)^{-1} dr^2 - r^2 d\Omega^2 + \bigg(1-\frac{r^2}{a^2}\bigg) dt^2.
\end{equation}
For a region very small compared to $a$, i.e. for $r\ll a$, this line element has Minkowski form
\begin{equation}\label{backgroundmetric}
d\sigma^2 = -dr^2 - r^2 d\Omega^2 + dt^2.
\end{equation}
The convention used here for coordinates is
$$x^1=r, x^2=\theta_1, x^3=\theta_2,\cdot\cdot\cdot, x^{n+1}=\theta_n, x^{n+2}=t. $$
The energy momentum tensor for charged anisotropic fluid is of the form
\begin{equation}\label{EMT}
T_{\mu \nu} = (\rho+p_\perp) U_\mu U_\nu - p_\perp g_{\mu \nu} + (p_r-p_\perp)\chi_\mu \chi_\nu + \frac{1}{4\pi}\bigg(g^{\lambda \alpha} F_{\mu \lambda} F_{\nu \alpha} - \frac{1}{4} g_{\mu \nu} F_{\lambda \alpha} F^{\lambda \alpha}\bigg),
\end{equation}
with matter density $\rho$, $p_r$ being the radial pressure in the direction of $\chi_\mu$, $p_\perp$ being the tangential pressure orthogonal to $\chi_\mu$, the $(n+2)$-velocity vector of the fluid $U_\mu$ and $\chi_\mu$ being the unit space-like vector orthogonal to $U_\mu$.\\
The skew symmetric Maxwell Tensor $F_{\mu \nu}$ satisfies the Maxwell's equations in the form
\begin{eqnarray}
F_{\mu \nu; \lambda} + F_{\nu \lambda; \mu} + F_{\lambda \mu; \nu} &=& 0 ,\\
{F^{\mu \nu}}_{;\nu} &=& -4 \pi J^\mu,
\end{eqnarray}
where $J^\mu = \sigma U^\mu$ is the $(n+2)$-current of the charge distribution with proper charge density $\sigma$ within the $n$-sphere. It is known that due to spherical symmetry the only non-vanishing components of $F_{\mu \nu}$ are $F_{1(n+2)}$ and $F_{(n+2)1}$.\\
Choosing the comoving system we write
\begin{eqnarray}
U^\mu &=& (0,0,0,\cdot\cdot\cdot (n+1) times, e^{-\frac{\nu}{2}}),\\
\chi^\mu &=&( e^{-\frac{\lambda}{2}},0,0,\cdot\cdot\cdot (n+1) times),
\end{eqnarray}
here $U_\mu U^\mu=-\chi_\mu \chi^\mu=1$.\\
The non-vanishing components of the energy momentum tensor are
\begin{eqnarray}
T_1^1 &=& -p_r +\frac{E^2}{8 \pi} , \\
T_2^2 = T_3^3 = T_4^4 = \cdot\cdot\cdot = T_{n+1}^{n+1} &=& -p_\perp - \frac{E^2}{8 \pi},\\
T_{n+2}^{n+2} &=& \rho+\frac{E^2}{8 \pi}.
\end{eqnarray}
In the region $r\ll a$ and neglecting the terms which are small throughout this region, we can write the non-vanishing components of $S_\mu^\nu$ (see \cite{FalikRosen1980}) as,
\begin{equation}\label{smunu}
-S_1^1=-S_2^2=-S_3^3=\cdot\cdot\cdot=-S_{n+1}^{n+1}=S_{n+2}^{n+2} =\frac{3}{2 a^2}e^{-\nu}.
\end{equation}
Using the procedure as given in \cite{Rosen1980}, the final form of field equations (\ref{EFE}) using Einstein tensor $G_{\nu}^{\mu}$ for metric (\ref{metric}), energy-momentum tensor (\ref{EMT}), background metric (\ref{backgroundmetric}) and values from equation (\ref{smunu}) are written as,
\begin{equation}\label{11component}
e^{-\lambda}\bigg[\frac{n \nu'}{2r}+ \frac{n(n-1)}{2 r^2}\bigg]-\frac{n(n-1)}{2 r^2} = 8 \pi p_r-E^2-\frac{3}{2a^2}e^{-\nu},
\end{equation}
\begin{equation}\label{22component}
e^{-\lambda}\bigg[\frac{\nu''}{2}- \frac{\lambda' \nu'}{4}+\frac{\nu'^2}{4}-\frac{(n-1)(\lambda'-\nu')}{2r}+\frac{(n-1)(n-2)}{2r^2}\bigg]-\frac{(n-1)(n-2)}{2r^2} = 8 \pi p_\perp + E^2-\frac{3}{2a^2}e^{-\nu},
\end{equation}
\begin{equation}\label{44component}
e^{-\lambda}\bigg[\frac{n\lambda'}{2r}-\frac{n(n-1)}{2r^2}\bigg]+\frac{n(n-1)}{2r^2} = 8 \pi \rho + E^2-\frac{3}{2a^2}e^{-\nu},
\end{equation}
\begin{equation}\label{extracomponent}
(r^n E)' = \frac{2 \pi^{\frac{n+1}{2}}}{\Gamma(\frac{n+1}{2})} r^n \sigma(r) e^{\frac{\lambda}{2}},
\end{equation}
Here a prime for $\lambda$ and $\nu$ denotes a differentiation with respect to r.\\
The energy-momentum conservation equation $T^\mu_{\nu;\mu}=0$ gives,
\begin{equation}\label{conservationofEM}
(\rho+p_r)\frac{\nu'}{2} + p_r' = \frac{n}{r}(p_\perp-p_r)+\frac{1}{8 \pi r^4}\frac{dQ^2}{dr}+\frac{(n-2)E^2}{4 \pi r}.
\end{equation}
where charge $Q$ is related with the field strength $E$, through the integral form of the Maxwell's equation (\ref{extracomponent}), which can be written as
\begin{equation}\label{Q(r)}
Q(r) = E r^n = 4\pi \int_0^r r^n e^{\lambda/2} \sigma(r) dr,
\end{equation}
We now define the \emph{effective} density $\rho_e$, \emph{effective} radial pressure $p_{r_e}$ and \emph{effective} tangential pressure $p_{\perp_e}$ (see \cite{HarpazRosen1985}) as,
\begin{eqnarray}\label{effective}
\begin{split}
\rho_e &=& \rho-\frac{3}{16 \pi a^2}e^{-\nu},\\
p_{r_e} &=& p_r-\frac{3}{16 \pi a^2}e^{-\nu},\\
p_{\perp_e} &=& p_\perp-\frac{3}{16 \pi a^2}e^{-\nu}.
\end{split}
\end{eqnarray}
So the field equations (\ref{11component})-(\ref{44component}) take the form,
\begin{equation}\label{11component*}
e^{-\lambda}\bigg[\frac{n \nu'}{2r}+ \frac{n(n-1)}{2r^2}\bigg]-\frac{n(n-1)}{2r^2} = 8 \pi p_{r_e}-\frac{Q^2}{r^{2n}},
\end{equation}
\begin{equation}\label{22component*}
e^{-\lambda}\bigg[\frac{\nu''}{2}- \frac{\lambda' \nu'}{4}+\frac{\nu'^2}{4}-\frac{(n-1)(\lambda'-\nu')}{2r}+\frac{(n-1)(n-2)}{2r^2}\bigg]-\frac{(n-1)(n-2)}{2r^2} = 8 \pi p_{\perp_e}+\frac{Q^2}{r^{2n}},
\end{equation}
\begin{equation}\label{44component*}
e^{-\lambda}\bigg[\frac{n\lambda'}{2r}-\frac{n(n-1)}{2r^2}\bigg]+\frac{n(n-1)}{2r^2} = 8 \pi \rho_e+\frac{Q^2}{r^{2n}}.
\end{equation}
Equation (\ref{conservationofEM}) can be rewritten as,
\begin{equation}\label{conservationofEM*}
(\rho_e+p_{r_e})\frac{\nu'}{2} + p_{r_e}' = \frac{n}{r}(p_{\perp_e}-p_{r_e})+\frac{1}{8 \pi r^4}\frac{dQ^2}{dr}+\frac{(n-2)Q^2}{4 \pi r^{2n+1}}.
\end{equation}
Now from equation (\ref{44component*}),
\begin{equation}\label{elambda}
e^{-\lambda}=1-\frac{2m_e(r)}{r}+\frac{2}{n(n-1)}\frac{Q^2}{r^{2n-2}},
\end{equation}
where $m_e(r)$ is the \emph{effective} mass function defined as,
\begin{equation}\label{m_e}
m_e(r)=\frac{1}{nr^{n-2}}\int_0^r \bigg(8 \pi \rho_e r^n+\frac{2QQ'}{(n-1)r^{n-1}}\bigg) dr.
\end{equation}
Now from equation (\ref{conservationofEM*}),
\begin{equation}\label{nudash}
\nu' = - \frac{2 p_{r_e}'}{(\rho_e+p_{r_e})} + \frac{2n(p_{\perp_e}-p_{r_e})}{r(\rho_e+p_{r_e})}+\frac{QQ'}{2 \pi r^4(\rho_e+p_{r_e})}+\frac{(n-2)Q^2}{2 \pi r^{2n+1}(\rho_e+p_{r_e})}.
\end{equation}
Using equations (\ref{elambda}) and (\ref{nudash}) in (\ref{11component*}), one can get
\begin{eqnarray}\label{11component**}
\bigg[1-\frac{2m_e}{r} &+& \frac{2}{n(n-1)}\frac{Q^2}{r^{2n-2}}\bigg]\bigg[-\frac{n r p_{r_e}'}{(\rho_e+p_{r_e})} + \frac{n^2(p_{\perp_e}-p_{r_e})}{(\rho_e+p_{r_e})}+\frac{n QQ'}{4 \pi r^3(\rho_e+p_{r_e})} \nonumber \\ &+& \frac{n(n-2)Q^2}{4 \pi r^{2n}(\rho_e+p_{r_e})}+\frac{n(n-1)}{2}\bigg] = 8 \pi p_{r_e}r^2 + \frac{n(n-1)}{2}-\frac{Q^2}{r^{2n-2}}.
\end{eqnarray}
Define a \emph{generating function} $G(r)$ as,
\begin{equation}
G(r) = \frac{1-\frac{2m_e}{r}+\frac{2}{n(n-1)}\frac{Q^2}{r^{2n-2}}}{8 \pi p_{r_e}r^2 + \frac{n(n-1)}{2}-\frac{Q^2}{r^{2n-2}}}\label{G(r)},
\end{equation}
and introduce the \emph{anisotropic function} $w(r)$ as,
\begin{equation}
w(r) = \frac{n^2(p_{r_e}-p_{\perp_e})}{(\rho_e+p_{r_e})}G(r)\label{w(r)}.
\end{equation}
Using equations (\ref{G(r)}) and (\ref{w(r)}) in equation (\ref{11component**}), we get
\begin{equation}\label{eqn3}
8\pi (\rho_e+p_{r_e})=\frac{-8 \pi n r p_{r_e}' G + \frac{2nQQ'G}{r^3} + \frac{2 n (n-2) Q^2 G}{r^{2n}}}{(1-\frac{n(n-1)}{2}G+w)}.
\end{equation}
Differentiation of equation (\ref{elambda}) gives,
\begin{equation}\label{elambdadiff}
e^{-\lambda} \lambda' = \frac{2m_e'}{r}-\frac{2m_e}{r^2}-\frac{4QQ'}{n(n-1)r^{2n-2}}+\frac{4Q^2}{nr^{2n-1}}.
\end{equation}
Adding $8\pi p_{r_e}$ on both sides of equation (\ref{44component*}) and then using equations (\ref{elambdadiff}), (\ref{elambda}) and (\ref{G(r)}), we get
\begin{equation}\label{eqn1}
8\pi (\rho_e+p_{r_e}) = \frac{n m_e'}{r^2}-\frac{n m_e}{r^3}-\frac{2QQ'}{(n-1)r^{2n-1}}+\frac{2Q^2}{r^{2n}}+\bigg(\frac{n(n-1)}{2r^2}+8\pi p_{r_e}-\frac{Q^2}{r^{2n}}\bigg)\bigg(1-\frac{n(n-1)}{2}G\bigg).
\end{equation}
Differentiation of equation (\ref{G(r)}) gives,
\begin{eqnarray}\label{eqn2}
\frac{nm_e'}{r^2}-\frac{nm_e}{r^3}-\frac{2QQ'}{(n-1)r^{2n-1}}&+&\frac{2Q^2}{r^{2n}}=-\frac{nG'}{2r}\bigg(8 \pi p_{r_e}r^2 + \frac{n(n-1)}{2}-\frac{Q^2}{r^{2n-2}}\bigg) \nonumber \\
&-& 4 \pi n r p_{r_e}' G - 8 \pi n p_{r_e}G + \frac{nQQ'G}{r^{2n-1}}-\frac{n(n-1)Q^2G}{r^{2n}}.
\end{eqnarray}
Using equations (\ref{eqn2}) and (\ref{eqn3}) into equation (\ref{eqn1}), we get
\begin{eqnarray}
&& 8\pi p_{r_e}'+\frac{(2-n(n+1)G-nr G')(1-\frac{n(n-1)}{2}G+w)}{nr G(1+\frac{n(n-1)}{2}G-w)}8\pi p_{r_e}\nonumber\\ && +\frac{(n-1)(2-n(n-1)G-nr G')(1-\frac{n(n-1)}{2}G+w)}{2r^3 G(1+\frac{n(n-1)}{2}G-w)} \nonumber\\ &&-\frac{(2+n(n-1)G-nrG')(1-\frac{n(n-1)}{2}G+w)}{nrG(1+\frac{n(n-1)}{2}G-w)}\frac{Q^2}{r^{2n}} +\frac{(1-\frac{n(n-1)}{2}G+w)}{(1+\frac{n(n-1)}{2}G-w)}\frac{2QQ'}{r^{2n}}\nonumber\\ &&-\frac{4QQ'}{r^4(1+\frac{n(n-1)}{2}G-w)}-\frac{4(n-2)}{(1+\frac{n(n-1)}{2}G-w)}\frac{Q^2}{r^{2n+1}}=0,
\end{eqnarray}
which is a linear equation in $p_{r_e}$. We obtain its solution as,
\begin{equation}\label{soln}
8\pi p_{r_e}=e^{-\int B(r) dr}[p_0+\int C(r)e^{\int B(r) dr} dr],
\end{equation}
where $p_0$ is a constant of integration,
\begin{eqnarray}\label{BandC}
&B(r)& = \frac{(2-n(n+1)G-nr G')(1-\frac{n(n-1)}{2}G+w)}{nr G(1+\frac{n(n-1)}{2}G-w)}, \nonumber \\
&C(r)& = -\frac{(n-1)(2-n(n-1)G-nr G')(1-\frac{n(n-1)}{2}G+w)}{2r^3 G(1+\frac{n(n-1)}{2}G-w)} \nonumber\\
&& +\frac{(2+n(n-1)G-nrG')(1-\frac{n(n-1)}{2}G+w)}{nrG(1+\frac{n(n-1)}{2}G-w)}\frac{Q^2}{r^{2n}} -\frac{(1-\frac{n(n-1)}{2}G+w)}{(1+\frac{n(n-1)}{2}G-w)}\frac{2QQ'}{r^{2n}}\nonumber\\ &&+\frac{4QQ'}{r^4(1+\frac{n(n-1)}{2}G-w)}+\frac{4(n-2)}{(1+\frac{n(n-1)}{2}G-w)}\frac{Q^2}{r^{2n+1}}.
\end{eqnarray}
From equation (\ref{m_e}),
\begin{equation}\label{m_e'}
\frac{m_e'}{r}=\frac{8 \pi \rho_e r}{n}+\frac{2QQ'}{n(n-1)r^{2n-2}}.
\end{equation}
Using equation (\ref{m_e'}) in (\ref{eqn2}) and then using equation (\ref{G(r)}), we get
\begin{eqnarray}\label{8pirhoe}
8 \pi \rho_e &=& \bigg(1-\frac{n(n-1)}{2}G\bigg)\frac{n}{2r^2}-\bigg(\frac{2(2n-3)}{n(n-1)}+(2n-3)G\bigg)\frac{nQ^2}{2r^{2n}}\nonumber \\&-&4 n \pi G \bigg(3 p_{r_e}+r p_{r_e}'-\frac{QQ'}{4 \pi r^{2n-1}}\bigg)- \frac{nr G'}{2} \bigg(8\pi p_{r_e} + \frac{n(n-1)}{2r^2}-\frac{Q^2}{r^{2n}}\bigg),
\end{eqnarray}
which is the expression for effective density $\rho_e$. Equation (\ref{w(r)}) gives,
\begin{equation}\label{s}
p_{\perp_e}=p_{r_e}-\frac{w(r)}{n^2 G}(\rho_e+p_{r_e}).
\end{equation}
Equations (\ref{elambda}) and (\ref{G(r)}) yields,
\begin{equation}\label{elambda*}
e^{-\lambda}=G\bigg(8\pi p_{r_e} r^2 + \frac{n(n-1)}{2}-\frac{Q^2}{r^{2n-2}}\bigg).
\end{equation}
Using this above equation in equation (\ref{11component*}), we obtain
\begin{equation}
\frac{d\nu}{dr}=\frac{2}{nr G}-\frac{(n-1)}{r},
\end{equation}
which after integration gives,
\begin{equation}\label{enu}
e^{\nu}=\frac{A^2}{r^{(n-1)}}e^{\frac{2}{n}\int (1/rG)dr},
\end{equation}
where $A$ is the constant of integration.\\
Using values from equations (\ref{enu}) and (\ref{elambda}), the spacetime (\ref{metric}) becomes
\begin{equation}
ds^{2} = - \bigg[1-\frac{2m_e(r)}{r}+\frac{2}{n(n-1)}\frac{Q^2}{r^{2n-2}}\bigg]^{-1} dr^2 - r^2 (d\theta^2 + \sin^2\theta d\phi^2) + \frac{A^2}{r^{(n-1)}}e^{\frac{2}{n}\int (\frac{1}{rG})dr} dt^2.
\end{equation}
\section{Special case:}
The local flatness at the origin is required for a physically meaningful solution. Thus we assume the non divergent effective pressure at origin, as $r \rightarrow 0$, $\frac{m_e(r)}{r} \rightarrow 0$ and $\frac{Q^2}{r^{2n-2}}\rightarrow 0$ which results into $G(r) \rightarrow \frac{2}{n(n-1)}$. If $G(r)=\frac{2}{n(n-1)}, w(r)=0(i.e.p_r=p_\perp)$ and $Q(r)=0$, one obtains $\lambda = 0$.\\
By considering the charge density $\sigma$ to be constant, we can get $Q(r)\thicksim r^3$ from equation (\ref{Q(r)}). The appropriate junction condition at the surface $r=r_0$ yields
\begin{equation}
Q(r) = e(r/r_0)^3.
\end{equation}
If we denote $e/r_0^3=K$ then we can write,
\begin{equation}
Q(r)=kr^3.
\end{equation}
Further we define \emph{Generating function} and \emph{Anisotropic function} from equations (\ref{G(r)}) and (\ref{w(r)}) respectively as,
\begin{eqnarray}
G(r) &=& \frac{2}{n(n-1)}(1-\alpha r^2)\label{G(r)*},\\
w(r) &=& -\alpha r^2\label{w(r)*},
\end{eqnarray}
where $\alpha$ is a constant. This choice is physically appropriate because function $G(r)\thicksim \frac{2}{n(n-1)}$ as $r\thicksim 0$.\\
From equation (\ref{BandC}), $B(r)=0, C(r)=6 K^2 r + 2(n-2) k^2 r^{5-2n}$. So from equation (\ref{soln}),
\begin{equation}\label{pre}
p_{r_e}=\frac{p_0}{8\pi}+\frac{3K^2 r^2}{8 \pi}+\frac{(n-2)}{(3-n)}\frac{k^2}{8 \pi} r^{6-2n}.
\end{equation}
If constant $p_0=0$, then
\begin{equation}\label{pre1}
p_{r_e}=\frac{3K^2 r^2}{8 \pi}+\frac{(n-2)}{(3-n)}\frac{k^2}{8 \pi} r^{6-2n}.
\end{equation}
Hence from equation (\ref{effective}) the radial pressure is given by,
\begin{equation}
p_r=\frac{3}{16 \pi a^2}\bigg(\frac{(1-\alpha r^2)^{\frac{n-1}{2}}}{A^2}\bigg)+\frac{3K^2 r^2}{8 \pi}+\frac{(n-2)}{(3-n)}\frac{k^2}{8 \pi} r^{6-2n}.
\end{equation}
Also from equation (\ref{8pirhoe}) the effective density is obtained as,
\begin{equation}\label{rhoe}
\rho_e=\frac{3 n \alpha}{16 \pi} + \frac{21 \alpha K^2 r^4}{8 \pi (n-1)}-\frac{15 k^2 r^2}{8 \pi (n-1)} - \frac{(4n^2-32n+55)}{(n-1)(3-n)} \frac{\alpha k^2}{8 \pi} r^{8-2n} + \frac{(6n^2-37n+54)}{(n-1)(3-n)}\frac{k^2}{8 \pi} r^{6-2n},
\end{equation}
which gives the energy density using equation (\ref{effective}) as,
\begin{eqnarray}\label{rhoe1}
\rho &=&\frac{3}{16 \pi a^2}\bigg(\frac{(1-\alpha r^2)^{\frac{n-1}{2}}}{A^2}\bigg)+\frac{3 n \alpha}{16 \pi} + \frac{21 \alpha K^2 r^4}{8 \pi (n-1)}-\frac{15 k^2 r^2}{8 \pi (n-1)} \nonumber \\
&-& \frac{(4n^2-32n+55)}{(n-1)(3-n)} \frac{\alpha k^2}{8 \pi} r^{8-2n} + \frac{(6n^2-37n+54)}{(n-1)(3-n)}\frac{k^2}{8 \pi} r^{6-2n}.
\end{eqnarray}
Using values from equations (\ref{G(r)*}), (\ref{w(r)*}), (\ref{pre1}) and (\ref{rhoe}) into equation (\ref{s}), one can obtain
\begin{eqnarray}
p_{\perp_e} &=& \frac{3 K^2 r^2}{8 \pi}+\frac{3(n-1)\alpha^2 r^2}{32 \pi (1-\alpha r^2)}+\frac{3(n-6)\alpha k^2 r^4}{16 \pi n(1-\alpha r^2)} + \frac{21 \alpha^2 k^2 r^6}{16 \pi n(1-\alpha r^2)}\nonumber \\ &-&\frac{(4n^2-32n+55)\alpha^2 k^2 r^{10-2n}}{16 \pi n(3-n)(1-\alpha r^2)}+ \frac{k^2 r^{6-2n}}{8 \pi (3-n)} \bigg\{\frac{(7n^2-40n+56)\alpha  r^2}{2 n(1-\alpha r^2)}+(n-2)\bigg\}.
\end{eqnarray}
Hence from equation (\ref{effective}) the tangential pressure is given by,
\begin{eqnarray}
p_{\perp} &=& \frac{3}{16 \pi a^2}\bigg(\frac{(1-\alpha r^2)^{\frac{n-1}{2}}}{A^2}\bigg) + \frac{3 K^2 r^2}{8 \pi}+\frac{3(n-1)\alpha^2 r^2}{32 \pi (1-\alpha r^2)}+\frac{3(n-6)\alpha k^2 r^4}{16 \pi n(1-\alpha r^2)} \nonumber \\ &+& \frac{21 \alpha^2 k^2 r^6}{16 \pi n(1-\alpha r^2)} -\frac{(4n^2-32n+55)\alpha^2 k^2 r^{10-2n}}{16 \pi n(3-n)(1-\alpha r^2)}\nonumber \\ &+& \frac{k^2 r^{6-2n}}{8 \pi (3-n)} \bigg\{\frac{(7n^2-40n+56)\alpha  r^2}{2 n(1-\alpha r^2)}+(n-2)\bigg\}.
\end{eqnarray}
Using equations (\ref{G(r)*}) and (\ref{pre1}), equation (\ref{elambda*}) can be written as
\begin{equation}\label{elambda**}
e^{-\lambda}=\frac{2}{n(n-1)}(1-\alpha r^2)\bigg(\frac{n(n-1)}{2}+3K^2 r^4+\frac{(2n-5)}{(3-n)}k^2 r^{8-2n}\bigg).
\end{equation}
Using equation (\ref{G(r)*}) into equation (\ref{enu}), we get
\begin{equation}\label{enu*}
e^{\nu}=\frac{A^2}{(1-\alpha r^2)^{\frac{n-1}{2}}}.
\end{equation}
Using values from equations (\ref{elambda**}) and (\ref{enu*}), the cosmological model for spacetime (\ref{metric}) is
\begin{eqnarray}\label{model}
ds^{2} &=& - \bigg[\frac{2}{n(n-1)}(1-\alpha r^2)\bigg(\frac{n(n-1)}{2}+3K^2 r^4+\frac{(2n-5)}{(3-n)}k^2 r^{8-2n}\bigg)\bigg]^{-1} dr^2 \nonumber \\ &-& r^2 (d\theta^2 + \sin^2\theta d\phi^2) + \frac{A^2}{(1-\alpha r^2)^{\frac{n-1}{2}}} dt^2.
\end{eqnarray}
\section{Conclusion:}
\goodbreak In this paper we have presented exact analytical solution of field equations of Bimetric General Relativity for the case of static spherically symmetric anisotropic distribution of charged matter in $(n+2)$-dimensions by introducing the \emph{generating function} and \emph{anisotropic function} as defined in (\ref{G(r)*}) and (\ref{w(r)*}). From equation (\ref{m_e}), we note that along with material density the electromagnetic anisotropy also contributes to the mass. It can also be noted that for $Q(r)=0$, the solution obtained here match with the solution of Kandalkar and Gawande (see \cite{kandalkarGawande2008}) for a neutral matter in Higher Dimensional General Relativity.
\goodbreak From the expression of $\rho$, $p_r$ and $p_{\perp}$ it is apparent that the last terms contribute negatively to these quantities. For large $n$, these terms go as reciprocal powers of $r$; thus for large $r$ they tend to zero.
\goodbreak Moreover the present result reduces to the Einstein's general relativity for a physical system which is small compare to the size of the universe because in such case the term $\frac{3}{2 a^2}e^{-\nu}$ in the field equations (\ref{11component})-(\ref{44component}) is negligible, this conclusion is derived by matching our results  with the one obtained by Singh et al. (see \cite{SinghHelmi1995}) for the 4-dimensional General Relativity. Moreover for $n=2$, results in this paper matches with the results obtained by Hasmani and Pandya (see \cite{HasmaniDisha}) for 4-dimensional anisotropic charged matter in BGR.\\

\noindent \textbf{Acknowledgement}
The authors are thankful to the University Grant Commission, India for providing financial support under UGC-SAP-DRS (III) provided to the Department of Mathematics, Sardar Patel University, Vallabh Vidyanagar, where the work was carried out. D.N. Pandya is also thankful to the UGC, India for providing financial support under UGC-BSR Fellowship (Grant No:F.4-1/2006 (BSR)/ 7-159/2007(BSR) dated 16/01/2014).

\end{document}